\documentclass{PoS}
\pdfoutput=1

\usepackage{graphicx}
\usepackage{amsmath,amsfonts}

\numberwithin{equation}{section}

\DeclareMathOperator{\Tr}{Tr}

\DeclareMathOperator{\SU}{SU}
\renewcommand{\phi}{\varphi}
\renewcommand{\epsilon}{\varepsilon}
\DeclareMathOperator{\C}{\cal C}

\newcommand{\R}{\mathbb R}

\title{Large-N string tension from rectangular Wilson loops}

\ShortTitle{Large-N string tension from rectangular Wilson loops}

\author{\speaker{Robert Lohmayer}\thanks{Research supported in part by the 
DOE, grant number DE-FG02-01ER41165.}
        \\
        Rutgers University, Department of Physics and Astronomy, Piscataway,
        NJ 08854, USA \\
        E-mail: \email{lohmayer@physics.rutgers.edu}}

\author{Herbert Neuberger\thanks{Research supported in part by the 
DOE, grant number DE-FG02-01ER41165.}\\
        Rutgers University, Department of Physics and Astronomy, Piscataway,
        NJ 08854, USA\\
        E-mail: \email{neuberg@physics.rutgers.edu}}

\abstract{In pure $\SU(N)$ gauge theory in four dimensions, we determine the string tension at large $N$ from smeared rectangular Wilson loops on the lattice. We learn how well loops of sizes barely on the strong-coupling side of the large-$N$ transition in their eigenvalue distribution can be described by effective string theory.}

\FullConference{The 30th International Symposium on Lattice Field Theory\\
		 June 24 - 29,  2012\\
		 Cairns, Australia}

\begin{document}

\section{Introduction}

We study Wilson loop operators $W(\C)$ in four-dimensional Euclidean $\SU(N)$ pure gauge theory, where $\C$ is a rectangular curve in $\R^4$. 
Perimeter- and corner-divergences of $W$ are eliminated by a (continuous) smearing procedure \cite{Narayanan:2006rf, Lohmayer:2011si}, where the associated smearing parameter $s$ of dimension length squared introduces an effective thickness for the curve $\C$.

In the infinite-$N$ limit, the eigenvalue spectrum of the Wilson loop matrix exhibits a non-analyticity \cite{Narayanan:2006rf,Lohmayer:2011nq}, separating a weakly-coupled short-distance regime from a qualitatively different strongly-coupled long-distance regime. 
At the transition point, the gap around -1 in the eigenvalue spectrum just closes. While smaller loops are insensitive to the compactness of $\SU(N)$, the full group is explored for larger loops (a key ingredient for confinement).
The transition point, which depends on the shape of $\C$, provides a natural scale for matching perturbation theory to the long-distance description provided by effective string theory \cite{Aharony:2010cx}. Our goal is to determine, by numerical lattice gauge theory methods, how well loops of sizes barely on the strong-coupling side of the large-$N$ transition can be described by effective string theory.

Using the standard single-plaquette Wilson action, we have obtained Monte Carlo estimates for smeared rectangular Wilson loops on a hypercubic lattice for various $N$'s, couplings, volumes, and loop sizes from a database of 160 uncorrelated equilibrated gauge fields. All statistical errors quoted below are determined by jackknife with the elimination of one single gauge configuration from the set of 160 at a time.
The range of couplings we use is $0.359\leq b \leq 0.369$, spaced by $\Delta b = 0.001$ (the upper bound prevents spontaneous $Z^4(N)$ symmetry breaking \cite{Kiskis:2003rd} on all our volumes $V\geq 12^4$). Satisfactory statistical 
independence for our observables is obtained for gauge fields at neighboring
$b$'s being separated by 500 complete $\SU(2)$ updates combined with
500 complete over-relaxation passes. The $N\to\infty$ limit is taken at fixed $b=\frac{\beta}{2N^2}$. The set of $N$-values we use consists of $N=7$, 11, 13, 19, 29. For the continuous smearing parameter, denoted by $S$ on the lattice, we mainly use $0.2\leq S \leq 0.4$.
The Wilson loops $W_N$ on the lattice are defined by 
\begin{align}
W_N (L_1,L_2,b,S,V)=\frac 1N \langle \Tr  \prod_{{\it l}\in \C} U_{{\it l}} \rangle\,.
\end{align}
The product is over the links ${\it l}$ in the order they appear when
one goes once round $\C$, a rectangle of sides $L_{1,2}$.   All our
fits are applied to 
\begin{align}
w_N(L_1,L_2,b,S,V)=-\log W_N(L_1,L_2,b,S,V) \,.
\end{align}
When the loops are square, the two variables $L_1,L_2$ are replaced
by one $L$ with the understanding that $L_{1,2}=L$.

In Sec.~\ref{sec:StringTension}, we present our results for the large-$N$ string tension obtained exclusively from square loops. In Sec.~\ref{sec:ShapeDependence}, we extract a purely loop-shape dependent number from the data and compare it to the prediction made by effective string theory. 
For a more detailed presentation and discussion we refer to Ref.~\cite{Lohmayer:2012ue}.

\section{String tension from square loops}
\label{sec:StringTension}
We first want to determine
$
\lim_{N\to\infty}\left (\lim_{V\to\infty} w_N(L,b,S,V)\right )
$
for square $L\times L$ loops.
In principle, large-$N$ reduction provides a shortcut for the infinite-volume limit. However, this requires tests and fits since finite-volume effects depend on $V$, $N$, $b$ and $L$.
We use two different methods to compute the limit:
\begin{itemize}
\item Method 1) 
\newline
At fixed $N$, we compute $w_N$ on volumes that are sufficiently large for
finite-volume effects to be negligible, then we determine
$w_{\infty}(V=\infty)$ (other arguments are omitted) by 
fitting $w_N(V=\infty)$ to 
\begin{align}\label{eq:infNinfVfit}
w_N(V=\infty)=w_\infty(V=\infty)+\frac{a_1(V=\infty)}{N^2}+\frac{a_2(V=\infty)}{N^4}\,.
\end{align}
We have evidence (strong for $N=7$ and $N=11$, not that strong for
  $N=19$ and rather weak for $N=29$) 
  that volumes $V=24^4$, $18^4$, $14^4$, $12^4$ 
  are sufficiently large for $N=7$, $11$, $19$, $29$, respectively.  
This statement applies to the specific set of couplings and loop sizes we use. 
\item Method 2) 
\newline
The second method makes use of large-$N$ reduction. At fixed $V$, we first
take the limit $N\to\infty$ of $w_N(V)$ by fitting
\begin{align}\label{eq:infNfixedVfit}
w_N(V)=w_\infty(V)+\frac{a_1(V)}{N^2}+\frac{a_2(V)}{N^4} \,.
\end{align}
So long as the center symmetry stays unbroken, 
there is no volume dependence in the infinite-$N$ theory, i.e.,
$w_\infty(V)=w_\infty(V=\infty)$.
We determine $w_\infty(V=12^4)$ 
from $N=11$, $13$, $19$, $29$ [method 2a)] and
$w_\infty(V=14^4)$ from $N=7$, $11$, $13$, $19$ [method 2b)].     
\end{itemize}
We obtain reasonable values of $\chi^2/N_{\text{dof}}$ for the fits\footnote{2a) is an exception where we have $\chi^2/N_{\text{dof}}$ up to 6 for $b\leq 0.361$. This probably reflects the 
  impact of the infinite-$N$ bulk transition at $b=0.360$.} 
 and good agreement (compatible with the statistical accuracy of
  about 0.1\%) between the three results for
  $\lim_{N,V\to\infty} w_N(V)$. 
Truncating the expansions \eqref{eq:infNinfVfit} and
  \eqref{eq:infNfixedVfit} at $\mathcal O(N^{-2})$ would result 
  in very large $\chi^2/N_{\text{dof}}$, so $a_2$ cannot be set to zero.
Including the $N=29$ result in the fit
  \eqref{eq:infNfixedVfit} for $V=12^4$ is crucial for 2a) to agree with 2b) and
  1) for large loops and large $b$. Including $N=7$ in the $V=12^4$ fit 
  would require an additional $1/N^6$
  correction in \eqref{eq:infNfixedVfit}.  
When the lattice size $V^{\frac 14}$ is getting close to the
critical lattice size $L_c(b)$ at which the center symmetry brakes, 
there is no useful information 
  to be gained about the $N,V=\infty$ limit 
  from numbers obtained at low values of $N$.
Since the required computation time scales as $N^3V$, 2a) is about
1.75 times more expensive than 2b) and 1) is about 2.5 times more expensive
than 2b). However, it is hardly possible to conclude from 2b) or 2a) alone that the estimates
for $w_\infty(V)$ are reliable. We became confident that we have correctly determined
$\lim_{N,V\to\infty} w_N(V)$ only after having obtained agreeing results
from 1), 2a) and 2b).

For fixed $b$ and $S$, we use the shorthand notation $w_\infty(L)\equiv\lim_{N,V\to\infty} w_N(L,b,S,V)$ and expect
\begin{align}\label{eq:wLL}
w_\infty(L)+\frac 14 \log L^2 = c_1 + c_2 L + \sigma L^2 + \mathcal
O\left(\frac 1{\sigma L^2}\right)\,.
\end{align}
The log term comes from the determinant of small fluctuations around 
the minimal area configuration in the effective string description. 
We shall return to it in Sec.~\ref{sec:ShapeDependence}. For now, its presence is just assumed 
because including it gives good fits while excluding it gives bad fits.

\begin{figure}[htb]
\begin{center}
\includegraphics[width=0.65\textwidth]{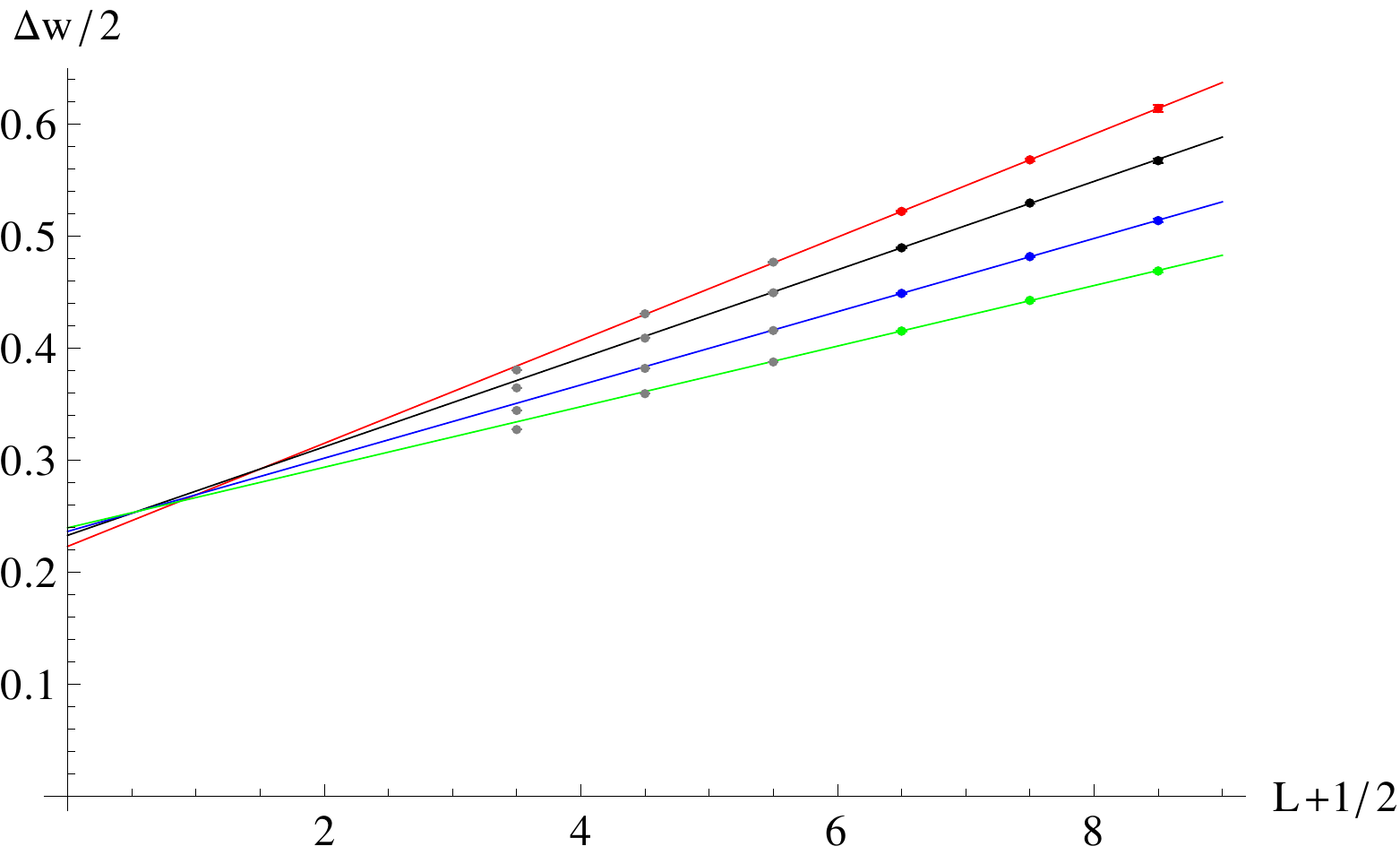}
\caption{Plots of $\frac{\Delta w}2=\frac 12 \left(w_\infty(L+1)-w_\infty(L)+\frac 12
  \log\left(1+\frac 1 L\right)\right)$ obtained with method 1) as a function
  of $L+\frac 12$ at $S=0.4$ and $b=0.36$ (red), 
  $b=0.362$ (black), $b=0.365$ (blue) and $b=0.368$ (green). Error bars are not visible in the plot. 
The straight lines show linear fits through the corresponding data points.
Only points $6 < L+\frac 12 < 9$ are used in the fits.} 
\label{fig:SigmaExample} 
\end{center}
\end{figure}

Neglecting corrections of order $\frac 1{\sigma L^3}$, we fit
\begin{align}\label{eq:SigmaFromSquares}
\frac 12 \left(w_\infty(L+1)-w_\infty(L)+\frac 12 
\log\left(1+\frac 1 L\right)\right)=\sigma \left(L+\frac 12\right)+\frac{c_2}{2}+ \mathcal
O\left(\frac 1{\sigma L^3}\right)
\end{align}
to a straight line as a function of $L+\frac 12$ to determine the string tension $\sigma$ and the coefficient of the perimeter term $c_2$ (see Fig.~\ref{fig:SigmaExample} for some examples). 
Most  $5\times 5$ loops fall
into the neighborhood of the large-$N$ phase transition in the
eigenvalue spectrum of the Wilson loop matrix for the $b$ and $S$ values we work with. 
Physically smaller loops will have a single-eigenvalue distribution which
has a gap around -1.
Therefore, we only use loop sizes in the range $6\leq L \leq 9$ to determine the string tension.

\begin{figure}[htb]
\begin{center}
\includegraphics[width=0.65\textwidth]{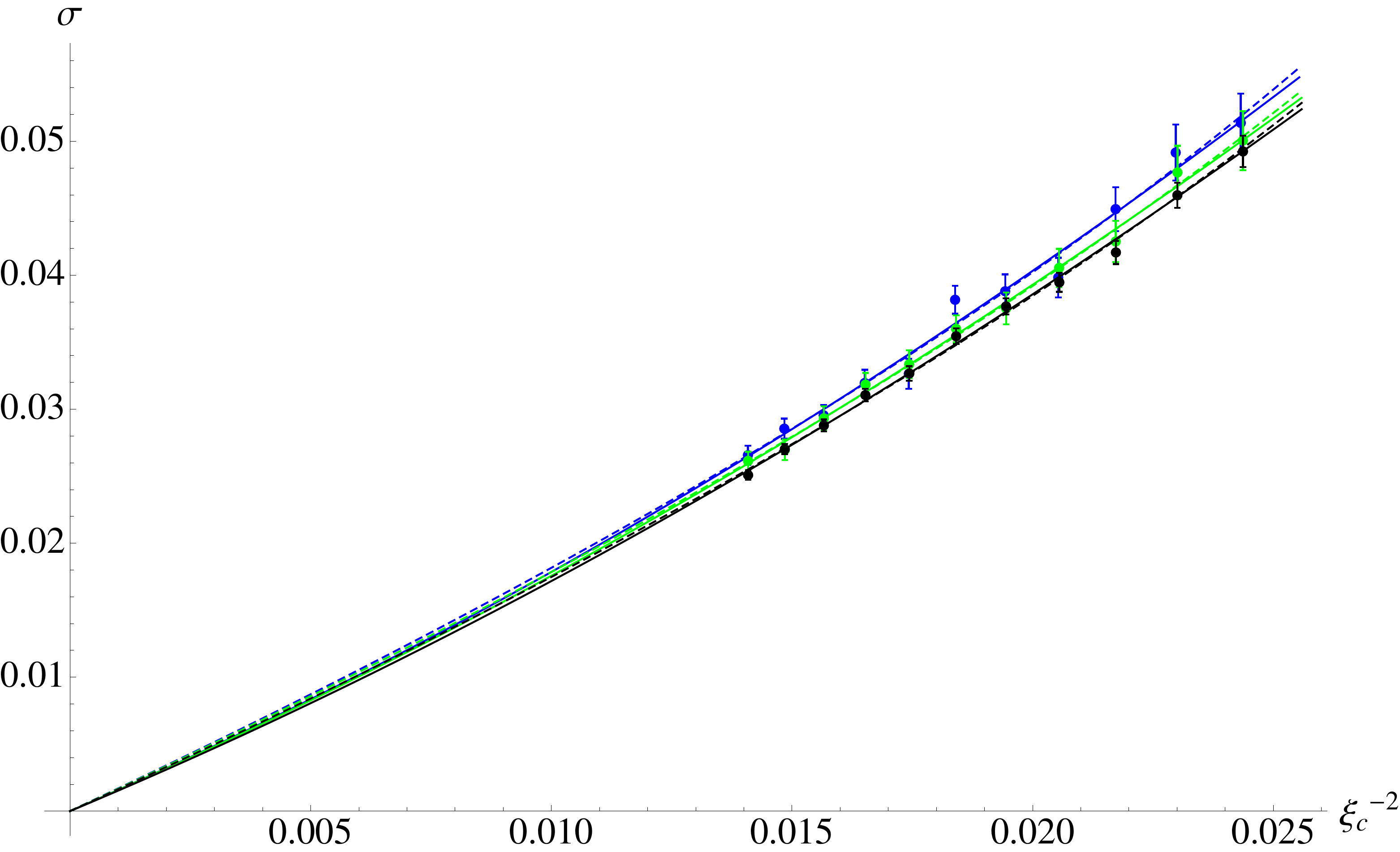}
\caption{
Plots of $\sigma$ as a function of $\xi_c^{-2}$: 
method 1) in black, method 2a) in green, method 2b) in blue, 
together with corresponding fit functions (the fits are obtained using $0.359\leq b \leq 0.369$).
} 
\label{fig:contfit1} 
\end{center}
\end{figure}

The results obtained for $\sigma$ using the
different methods for computing $w_\infty$ agree with each other within statistical errors, which are smallest for method 1). See Fig.~\ref{fig:contfit1} for a plot. As expected, $\sigma$ does not depend on the smearing level $S$ within the errors (which increase with decreasing $S$).

A scale length in lattice 
units denoted by $\xi_c(b)$ is used to carry out extrapolations to the continuum
limit. It is defined at $N=\infty$ using a three-loop calculation of 
the $\beta$-function for the lattice coupling. 
The coefficients are written as 
$\bar \beta_0=\frac{\beta_0}{N}=\frac{11}{48 \pi^2}$, 
$\bar \beta_1=\frac{\beta_1}{N^2}=\frac{34}{3 (16 \pi^2)^2}$,
 and $\bar \beta_2=\lim_{N\to\infty} \frac{\beta_2}{N^3}\approx -3.12211\times 10^{-5}$.
Integrating the RG flow, we define:
\begin{align}\label{eq:Lc}
\xi_c(b)=0.26 \left(\frac{\bar \beta_1}{\bar \beta_0^2}+\frac{b_I(b)}{\bar
  \beta_0}\right)^{-\frac{\bar \beta_1}{2\bar\beta_0^2}}\, \exp\left[\frac{b_I(b)}{2\bar
  \beta_0}\right]\, \exp\left[\frac{\bar \beta_2}{2\bar\beta_0^2 b_I(b)}\right]\,.
\end{align}
Above we have replaced the gauge coupling $b$ by
$b_I(b)=\lim_{N,V\to\infty} b\, W_N(L=1,b,S=0,V)$,
 a substitution known
as tadpole improvement. 
The definition of $\xi_c(b)$ is taken to match with \cite{Allton:2008ty}. 
We only added a numerical prefactor to make  $\xi_c(b)\approx L_c(b)$, where $L_c(b)$ 
is given in~\cite{Kiskis:2003rd}\footnote{ 
This approximation is good to 10-15\% in our range of couplings and would 
become exact at $b=\infty$.}. 

We separately carry out
two two-parameter fits of the relation between the string tension $\sigma(b)$ 
and $\xi_c(b)$. The two pairs of parameters are denoted by 
$d_0$, $d_1$ and $f_0$, $f_1$:
\begin{align}\label{eq:continuum}
\sigma(b)=\frac{d_0}{\xi_c(b)^2}+\frac{d_1}{\xi_c(b)^4}
\,,\qquad
\text{and} 
\qquad
\frac1{\xi_c(b)^2}=f_0^{-1} \sigma(b)+f_1 \sigma(b)^2\,.
\end{align}
We use ranges $0.359 \leq b \leq 0.369$ (range A) and $0.362 \leq b \leq 0.367$ (range
B). We also use the limited $b$ range (B) since we have observed increasing
$\chi^2$'s for the infinite-$N,V$ extrapolations using method 2a) for $b\leq
0.361$, as mentioned above. Another reason is that finite-volume
effects increase with increasing $b$. This reason only applies to method 1). 
 The difference between the two fits is a simple indicator of systematic errors
 induced by the truncation of the perturbative series. 
We find that these particular systematic deviations are of the same order as the  statistical errors. 

Our result for the infinite-$N$ continuum string tension 
is given by 
$
\lim_{b\to\infty}\sigma(b)\xi^2_c(b)=1.6(1)(3)
$.
The first error is statistical and the second systematic. The systematic error is more of a guess than a well founded estimate. In terms of $\Lambda_{\overline{MS}}$, this translates to $\sigma/\Lambda_{\overline{MS}}^2=3.4(2)(6)$.

Our central number is 2-3 standard deviations smaller than that of Allton et al.~\cite{Allton:2008ty} ($\sigma/\Lambda_{{\overline MS}^2} = 3.95(3)(64)$ at $N=\infty$).
Their numbers were extracted from Polyakov loops which are substantially
 longer than one side of our square loops. 
The systematic errors are dominated by the continuum extrapolation and 
 their relative size is roughly the same for us. 

A previous estimate for the string tension at infinite $N$ extracted from rectangular Wilson loops has been given in \cite{Kiskis:2009rf}.  Expressed in terms  of our variables it is $\sigma\xi_c^2\vert_{b_I=0.182} =2.2(3)$. 
While writing up our paper \cite{Lohmayer:2012ue} a new study~\cite{GonzalezArroyo:2012fx} appeared which also deals with
rectangular Wilson loops. These authors obtain $\sigma/\Lambda_{{\overline MS}^2} = 3.63(3)$ (statistical error) at $N=\infty$ if they apply the continuum extrapolation method of~\cite{Allton:2008ty}. 
This number is fully consistent with ours and has very small errors by
comparison. 

There seems to be a disagreement at the statistical level between~\cite{Allton:2008ty} and 
our result which agrees with that of~\cite{GonzalezArroyo:2012fx}. The result of~\cite{Kiskis:2009rf} seems to
side with that of~\cite{Allton:2008ty}, but has too large errors to be sure. 
The systematic errors are too large to claim evidence for a difference 
between the string tension extracted from Wilson loops and that extracted from
Polyakov loop correlators, which would be very difficult to accept at the theoretical level. 

\section{Shape dependence}
\label{sec:ShapeDependence}

We now turn to a 
study of the shape dependence of the scale-independent term 
in $w_N$ and compare it with the
effective-string prediction.  
For rectangular $L_1\times L_2$ loops 
it is convenient to introduce the
modular invariant shape parameter
\begin{align}
\zeta=\frac{L_1}{L_2}+\frac{L_2}{L_1}\,.
\end{align}  

The accuracy we now need does not permit taking the $N\to \infty$ limit.
We therefore restrict our attention to the $N=7,11$ data. We shall see that the numbers we
compute are identical within errors for $N=7$ and $N=11$, indicating that 
it is unlikely that they will change in a substantial manner in the $N=\infty$ limit. 

\begin{figure}
\begin{center}
\includegraphics[width=0.65\textwidth]{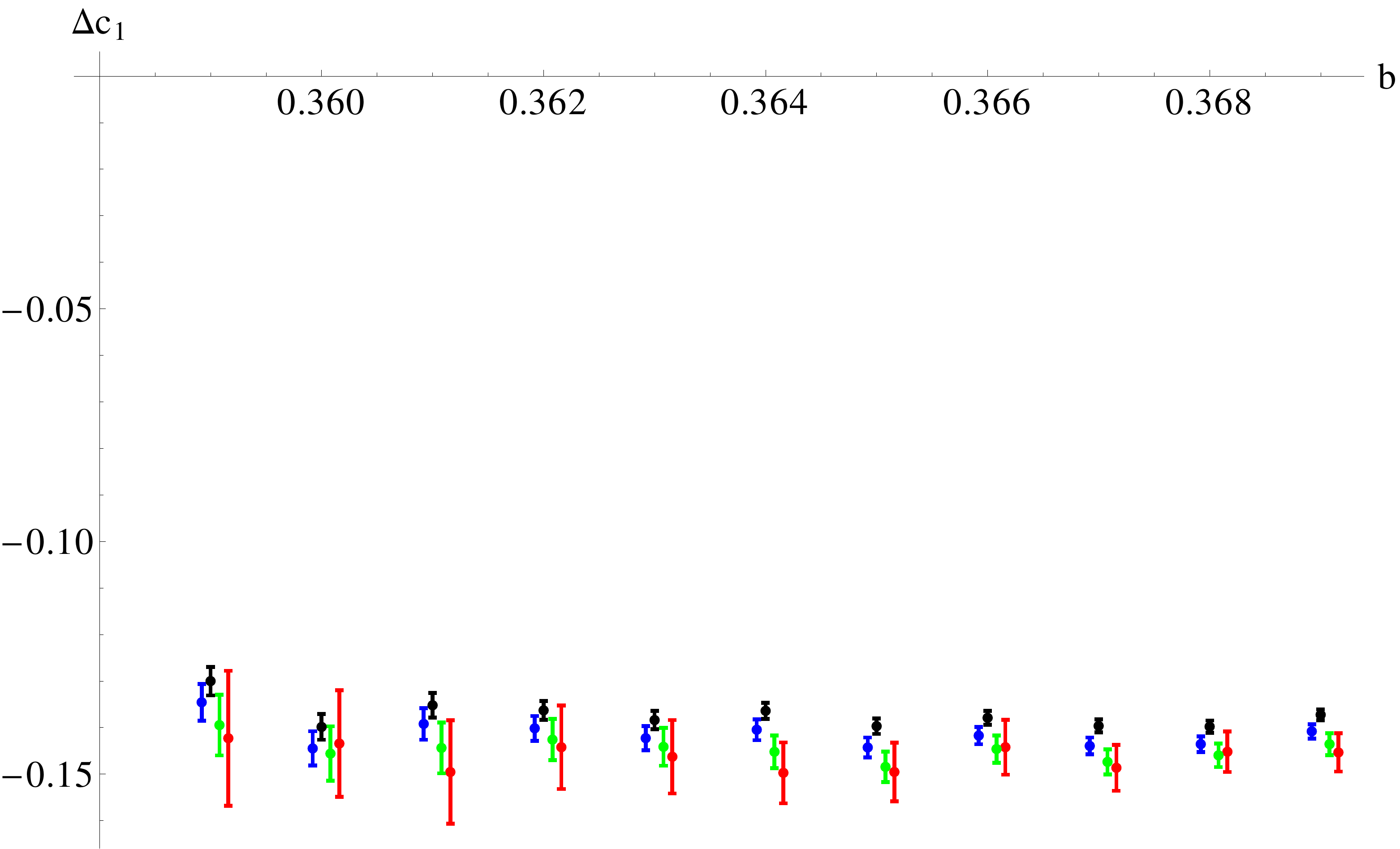}
\caption{Plot of $c_{1,N}(2.5)-c_{1,N}(2)$ for $N=11$ (on $V=18^4$) as a
  function of $b$ for $S=0.2$ (red), $S=0.28$ (green) $S=0.4$ (blue), and $S=0.52$ (black).} 
\label{fig:N11-DeltaC1} 
\end{center}
\end{figure}

At fixed $b$, $S$, $V$, and fixed
finite $N$, we expect (arguments $b$, $S$, $V$ are omitted)
\begin{align}\label{eq:wL1L2}
w_N(L_1,L_2)+\frac 14 \log L_1 L_2 = c_{1,N}(\zeta) + c_{2,N}
\frac{L_1+L_2}{2} + \sigma_N L_1 L_2 + \mathcal
O\left(\frac 1{\sigma_N L_1 L_2}\right)\,.
\end{align}
After having determined the lattice string tension $\sigma_N$ and the coefficient of the perimeter term $c_{2,N}$ from square $L\times L$ loops (using the method described in Sec.~\ref{sec:StringTension}) at fixed $N$, $b$, $S$, $V$, we fit $w_N(L,L)+\frac 14 \log L^2-\sigma_N
L^2 - c_{2,N} L$ to a constant, $c_{1,N}(\zeta=2)$ (using loop sizes $6\leq L \leq10$).
Next, we analyze the results obtained for 
a sequence of rectangular loops at the same $b$, $S$, $V$,
$N$ with $L_2=2 L_1$, i.e., $\zeta=\frac 52$ fixed. 
Using the results for $\sigma_N$ and $c_{2,N}$ obtained from square loops, we then determine $c_{1,N}(\zeta=\frac 52)$ by fitting
$w_N(L,2L)+\frac 14 \log\left( 2L^2\right)-\sigma_N
2L^2 - c_{2,N} \frac 32 L$ to a constant (using $4\leq L \leq 7$). 
Figure~\ref{fig:N11-DeltaC1} shows a plot of $c_{1,N=11}(2.5)-c_{1,N=11}(2)$ as a
function of $b$. Within statistical errors, our results for
$c_{1,N}(2.5)-c_{1,N}(2)$ do not depend on $b$, $S$, or $N$.
The effective-string prediction for $c_{1}(2.5)-c_{1}(2)$ is
\begin{align}
\frac 12 \log\left(\frac{\eta(2 i)\eta(i/2)}{\eta(i)^2}\right)\approx-0.08664\,,
\end{align}
where $\eta(x)$ is the eta-function. We find that the effective-string prediction 
is smaller than the observed 
values  by a factor of about 1.5 to 1.7.

We use sequences of $L\times L\pm1$ loops to cross check our results for the string tension and the shape dependence of $c_{1,N}$. 
Furthermore, we simultaneously fit $L\times L$, $L\times 2L$, and $L\times L\pm1$ loops to the functional form \eqref{eq:wL1L2}, where we expand $c_{1,N}$ around $\zeta=2$ and allow the coefficient of the $\log L_1L_2$ term to become a fit parameter. This analysis confirms the expected value of $1/4$ for the coefficient of the $\log$ term as well as the results for $\sigma_N$ and $c_{1,N}$ presented above. 

This means that one prediction coming from the 
determinant of Gaussian surface fluctuations in the effective string description works close to the 
large-$N$ transition in the eigenvalues and the other does not. However, these two predictions are somewhat different even within effective string theory (see \cite{Lohmayer:2012ue} for details).

Our result for the shape dependence of the scale-independent term is in agreement with \cite{GonzalezArroyo:2012fx} who independently report a deviation from effective string theory.

A detailed discussion of possible explanations (from perturbation theory and by higher-order terms in the string expansion) is presented in \cite{Lohmayer:2012ue}. Our data is not conclusive enough to settle this issue. We think that the shape dependence of planar Wilson loops presents an interesting case for testing the limitations of the effective string approach which deserves further study in the future.

\bibliographystyle{JHEP}
\bibliography{Lattice2012PoS}

\end{document}